\documentclass[a4paper]{jpconf}
\usepackage{graphicx}
\usepackage{amssymb}

\RequirePackage{color}

 \definecolor{MyDarkGreen}{rgb}{0.02,0.60,0.06}

\begin{document}
\title{Academic research groups: evaluation of their quality and quality of their evaluation}

\author{Bertrand~Berche$^{1,4}$, {Yurij}~Holovatch$^{2,4}$, Ralph~Kenna$^{3,4}$ and Olesya~Mryglod$^{2,4}$} 

\address{$^1$ Universit\'e de Lorraine, Statistical Physics Group, IJL, UMR CNRS 7198, Campus de Nancy, B.P. 70239, 54506 Vand\oe uvre l\`es Nancy Cedex, France}

\address{$^2$ Institute for Condensed Matter Physics of the National Academy of Sciences of Ukraine, 1 Svientsitskii Str., 79011 Lviv, Ukraine}

\address{$^3$ Applied Mathematics Research Centre, Coventry University, CV1 5FB, England}

\address{$^4$ The Doctoral College for the Statistical Physics of Complex Systems,
Leipzig-Lorraine-Lviv-Coventry $({\mathbb L}^4)$.}

\ead{bertrand.berche@univ-lorraine.fr; hol@icmp.lviv.ua; r.kenna@coventry.ac.uk; olesya@icmp.lviv.ua}

\begin{abstract}
In recent years,  evaluation of the quality of academic research has become an increasingly important and influential business. 
It determines, often to a large extent, the amount of research funding flowing into universities and similar institutes from governmental agencies and it impacts upon academic careers. 
Policy makers are becoming increasingly reliant upon, and influenced by, the outcomes of such evaluations. 
In response, university managers are increasingly attracted to simple {indicators} as guides to the dynamics of the positions of their various institutions in league tables. 
However, these league tables are {frequently} drawn up by inexpert bodies such as newspapers and magazines, using {rather} arbitrary measures and criteria.
Terms such as ``critical mass'' and {``metrics''} are {often} bandied about without {proper} understanding of what they actually mean. 
Rather than accepting the rise and fall of universities, departments and individuals on a turbulent sea of arbitrary measures, we suggest it is incumbent upon the scientific community itself to clarify their nature. 
Here we report on recent attempts to do that by  properly defining critical mass and showing how group size influences research quality. 
We also  examine currently predominant metrics and show that these fail as reliable indicators of group research quality. 
\end{abstract}

\section{Introduction}
\label{I}
\setcounter{equation}{0}

Academia is undergoing profound change.
All over the world, positions of leadership and senior management within universities are becoming  professionalised.  
Whereas in times gone by, such positions were typically occupied by senior academics, having proved themselves through long and successful careers in research and education, nowadays professional  managers -- who lack such experience --  are increasingly empowered. In the never-ending drive for increased efficiency, policy makers and university managers seek to commoditise all aspects of academic activity including research.
In this context, a plethora of indicators has emerged {{seeking}} to measure  the previously unmeasurable. 
As a result, the academic sector is becoming ever more  business-like and league tables  encourage competition over cooperation.

One of the simplest indicator is the size of a research department or group. 
Funding bodies, governments and universities frequently claim that these must be above a certain minimum  to be viable. 
The term ``critical mass'' has been borrowed from nuclear physics to capture this notion.
An implication of the term, as it is still all too widely understood, is that a group or department which is subcritical does not have the ability to sustain quality research activity. 
It suggests (erroneously as we shall see) the existence of a quantum jump in quality once the critical threshold is passed -- a first-order phase transition, to use the parlance of statistical physics. 
Extending the notion, it is often claimed that bigger is always better in research and that funding should therefore be focused into small numbers of large, elite institutions with smaller universities relegated to teaching-only roles.

Here we show that, despite their widespread uses,  these notions are manifestly incorrect. 
They are based upon unfounded and conflated analogies and not on scientific rigour.
A proper, measurable definition of critical mass in the research context has only  emerged in the last five years~\cite{KeBe10,KeBe11a,KeBe12a}.
It does {\emph{not}} entail a quantum leap or first-order transition from low  to high quality. 
Instead, the critical-mass phase transition is higher order -- barely noticeable, in fact.
There is also a second, measurable phase transition as group size grows. 
This is triggered by a {\emph{Ringelmann-type effect}}  and marked by a {\emph{Dunbar number.}}
The former is due to a slow-down in the rate of increase of quality with quantity due to coordination losses. 
The latter expresses the reasons for these losses, namely cognitive limits to the numbers of interactions individuals can sustain.
In Section~2 we explain these terms in more detail and review their connection to critical mass.

A second indicator which has recently emerged is the $h$-index. 
Again this is very attractive to policy makers and managers as a simple, zero-dimensional metric which is supposed to measure quality on an individual level. 
The notion has recently been extended to research groups in Ref.\cite{Bi14}.
Competing measures include the normalised citation index (NCI)~\cite{Evidenceweb,Ev10,Evidence2011}.
While the $h$-index itself was developed by a physicist, the NCI is an invention of a private company.
In the second part of this paper, we compare both of these against expert peer review measures of the quality of research groups. 
We show that the $h$-index is the better indicator of group quality but that its role is  strongly dependent on group size. 
We also show that neither of these metrics is a reliable substitute for expert peer review estimates of quality.

Academics fear the increased use of metrics to measure research activity.
The fear is that the increasingly professionalised management class lacks  detailed knowledge of, and appreciation for, academic subjects and seeks to base judgment and decisions on automated metrics rather than expertise and academic experience. 
This, accompanied by a top-down management approach, bolstered by metrics, and in pursuit of league-table rankings, may impinge upon  academic freedom as researchers are forced  to chase arbitrary measures rather than follow where their curiosity leads.
Here we show that these fears are well founded; despite the rising tide of metrics, they are a poor measure of quality.

\section{Evaluation of Quality and Critical Mass}
\label{II}
\setcounter{equation}{0}

The notion of critical mass in research has been around for a very long time.
The basic idea is that sub-critical research groups tend to produce  research of poor quality.
Once the critical threshold is passed, research quality becomes of an acceptable level.
This idea has been extended to, and far beyond, its logical conclusion to the idea that ``bigger is always better'' so that benefit always accrues through increasing group size. 
However, multiple analyses based on citation counts have produced no evidence for such a concept of critical mass.
Despite that, policies based upon this belief have been developed and implemented and these have serious consequences. 

Mark Harrison has studied how government officials  allocated 
funding amongst  research projects in the former USSR.
In Ref.\cite{Ha09}, he describes how policies swung between phases of competition and phases 
based on critical mass. 
Harrison sees remarkably similar parallels in the UK's funding system today.
He predicts that the current focus on concentrating research funding in pursuit of
 ``critical mass'' {{(which itself is misunderstood as we shall see)}} will eventually give way to competition, but only after the follies of the policy are declared.
By then, he fears, emerging research groups in promising universities will have lost their funding; small but excellent research centres will have closed and  individual careers curtailed.
Amazingly, the policies in which these dangers are inherent are based upon  concepts of critical mass which have no foundation.

Five years ago, we developed a theory for critical mass and the dependency of research quality on group size~\cite{KeBe10}. 
Our model was successfully tested against empirical data coming from peer-review based national research assessment exercises in the UK and France~\cite{KeBe11a}. 
To date, this model remains the only theory in existence for the relationship between research quality and group size. A significant amount of literature has been developed since Ref.\cite{KeBe10} and all quantitative evidence supports the theory. 
There is no quantitative evidence invalidating the model we are about to describe.
Here we   recall the model developed in Ref.\cite{KeBe10} and how it was statistically tested against empirical evidence.

It is important to realise that what we are about to develop here is a theory of averages (a ``mean-field theory'' in the language of statistical physics). 
There will always be deviations from such averages; the brilliant lone researcher, beavering away on a profound topic for years - still exists, especially in an area like pure mathematics~\cite{Nature}. 
(We will show, in fact, that critical mass for that subject is very small -- only one or two and perhaps it is unmeasurable.)
Also, although the model predicts higher research success for large groups than small, there are exceptions: some smaller groups are as good as, or better than some large ones.
We could overlay our model of averages with a random distribution to simulate deviations that always exist, but we feel such complications would only serve to obfuscate the trends we are trying to capture.
With this in mind, then, we proceed to develop our theory of averages.

A naive assumption (but one that was/is widespread, especially amongst policy makers, funding bodies and research managers) would be  that the strength $S$ of a research group or department is a simple sum of the strengths of the individuals comprising it.
The idea here is that excellent researchers will be attracted to larger research groups, and the best therefore become bigger. 
With this causation arrow, the driving mechanism is quality determining size: better becomes bigger.
If this were the only relation between quality and size, one would expect to see an unbounded {\emph{Matthew effect.}} 
The Matthew effect, or  cumulative advantage,  is a phenomenon whereby ``the rich get richer and the poor get poorer''.
(The name comes from the Gospel according to St Matthew, which states: ``For unto every one that hath shall be given, and he shall have abundance: but from him that hath not shall be taken away even that which he hath.'')
However, as we shall see, the empirical data does not support this scenario as the dominant driving mechanism.

Of course, individual strength is itself a  function of many factors, including 
innate calibre, teaching and administrative loads, management support,
the extent of interdisciplinarity, the availability of suitable equipment, whether the work is mainly
experimental, theoretical or computational, the methodologies and traditions of the field,
library facilities, journal access, external collaboration, grants held, confidence supplied by previous successes, prestige of the institute and many other factors.
We denote the mean strength of the individuals in a group, resulting from all these factors and more, by $a$.
If  the group has $N$ members, then, the strength of the group would simply be  $S=aN$.

However, research groups are complex systems and we have to take interactions between individuals  into account.  
We suppose that direct two-way interaction between individuals produces an added effect and we denote the mean strength of that effect by $b$. 
(We consider three-way interactions, and so on, as collections of two-way interactions, so we need not take these into account separately.) 
Since there are $N(N-1)/2$ possible two-way communication links in the group, the strength becomes $S= aN + bN(N-1)/2$.  
Note that certain institutions, publications and league tables refer to  ``research power'' rather than ``strength''~\cite{REF2014def}. 
We define research {\emph{quality}} $s$ as the strength or power per head, and write
$s = a_1 + b_1 N$. 

In fact this is not the full story for there has to be a limit to the number of colleagues with whom a given researcher can effectively communicate. 
In  {{evolutionary psychology and}} anthropology, this limit is known as the {\emph{Dunbar number}}~\cite{Du92}.
This was proposed in the 1990's by Robin Dunbar who noticed a correlation between the average  sizes of the neocortex in species of primates and the average sizes of their social groups.
Dunbar suggested that it represents a cognitive limit to the number of individuals  with whom  one can maintain stable  relationships. 
Extrapolating that correlation to human brain sizes, Dunbar predicts that we can comfortably maintain social relationships, on average, with about 150 people.  
Again,  this is an average only - the number can, of course, vary; it has been proposed to typically lie between 100 and 250.
There is a large body of evidence in support of Dunbar's theory:   
the estimated average size of a Neolithic farming village is 150, as is the fragmentation point  of Hutterite settlements. It is also   the size of basic units in professional armies in Roman antiquity as well as in  modern times since the 16th century. It is the average village size in the Domesday Book.

In our original publications on the topic, we referred to this limit as the upper critical mass and denoted it $N_c$. But it is a Dunbar number for academic researchers and it is discipline dependent.
 Once the group exceeds $N_c$, groups tends to fragment into subgroups. 
We suppose that the average number of such subgroups is  $N/(\alpha N_c)$, so that they have average size $\alpha N_c$. 
The strength of the group is then ascribed to the accumulation of individual strengths together with the strength intra-subgroups interactions,
$S= aN + bN(\alpha N_c-1)/2$.  
Of course, each of the $N/(\alpha N_c)$ subgroups can themselves interact with another subgroup. If the strength of such interactions is $c$, say, the total strength of an average large ($N > N_c$) group is expected to be 
$S = aN +  bN(\alpha N_c-1)/2 + c N/(\alpha N_c)[N/(\alpha N_c) - 1]/2$.
Gathering terms of the same order in $N$, we thus arrive at
\begin{equation}
 s = \left\{ \begin{array}{ll}
             a_1 + b_1 N &  {\mbox{if $N \le N_c$}} \\
             a_2 + b_2 N &  {\mbox{if $N \ge N_c$}},
             \end{array}
     \right.
\label{Nc}
\end{equation}
where $a_i$ and $b_i$ are functions of the strength parameters $a,b,c$ as well as $\alpha N_c$.
In particular $b_2 \propto 1/N_c^2$ so that slope of the $s$ versus $N$ curve to the right of the Dunbar point should be small for disciplines
with large $N_c$ values.

To test the above theory, we used empirical data from the  UK's last Research Assessment Exercise (RAE).
The RAE is a peer-review based assessment of the quality of university research which the Government then uses to decide how funds should be allocated between different universities, institutions and departments. 
Not surprisingly, it is taken very seriously  by university managers as well as by researchers themselves. 
The exercise took place about every five years up to 2008.
In 2014 it was replaced by a somewhat different exercise wherein, besides
quality of research, the impact which that research had beyond academia played an important role.
The 2014 event was called the Research Excellence Framework or REF.

While for the RAE  research was categorised into 67 academic disciplines, in the 2014 REF there were only 36 units of assessment (UoA). 
Applied Mathematics, for example, which includes some theoretical physics, was a stand-alone  UoA at RAE2008 but it was merged with Pure Mathematics, Statistics and Operational Research to form a Mathematical Sciences UoA for REF2014. 
The 2008 {exercise} was therefore a more ``fine-grained'' one and for that reason the data we report on here comes from the RAE.

In the UK system, research was scrutinised by experts from each discipline to determine the proportions carried out at each of five levels:
4* (world-leading research);
3* (internationally excellent research);
2* (research that is internationally recognised); 
1* (research recognised at a national level) and unclassified research.
Following RAE, a formula is used to determine how funding is distributed to higher education institutes for the subsequent years. The original formula used by the Higher Education Funding Council for England (HEFCE) valued 4* and 3* research seven and three times more highly than 2* research and allocated no funding to lower quality research. 
We can use that formula for the relative values of the various categories of research as a proxy for quality.
If $p_{n*}$ represents the percentage of  a team's
research which was rated  $n$*, then, the team's 
{\emph{quality} is
\begin{equation}
 s = p_{4*} + \frac{3}{7}p_{3*} + \frac{1}{7}p_{2*}
\,.
\label{seven}
\end{equation}

\begin{figure*}[t]
\begin{center}
\includegraphics[width=0.47\columnwidth,angle=0]{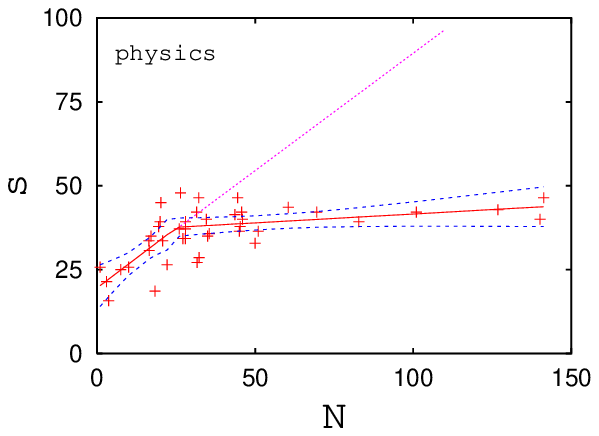}
\includegraphics[width=0.47\columnwidth,angle=0]{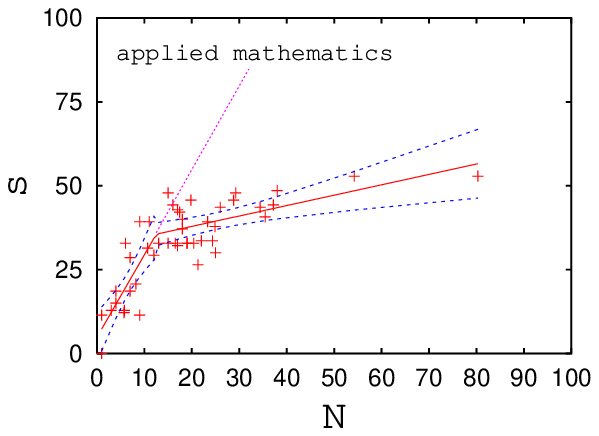}
\label{figJ6}
\end{center}
\vspace{0.3cm}
\caption{The relationship between quality and quantity of research groups submitted to the UK's last Research Assessment Exercise in Physics and Applied Mathematics. 
For Physics (which is dominated by experimental physics), the Dunbar number marking the breakpoint in the fitted red curve is $N_c = 25 \pm 5$. For Applied Mathematics (which includes some theoretical physics.) it is $13\pm 2$.
The critical mass is half the  Dunbar number. The blue dashed lines {in the plots} are error boundaries. The purple dotted lines are  extensions of the sub-breakpoint fits and capture some of the top performers.}
\end{figure*}

With Eq.(\ref{seven}) to hand, it is a straightforward exercise to translate the RAE results to quality scores and plot them against group size.\footnote{While generally throughout this paper, the word `group' means a collection of researchers at a given university active in a common
discipline, in the UK context it means those who were submitted to RAE2008 in a given unit of assessment. Usually these are permanent staff but they may include postdocs. A group is not, therefore, synonymous with a department because not all department members may be research active and, therefore, not all may have been submitted to RAE.
Alternatively, a submission may draw from researchers interacting across different departments. The word `group' in this sense is also not synonymous with research centre, as such entities may be involved more than one department.  Indeed, at RAE,
research centres may have been involved in submissions across more than one subject area.
While ``group'' in this context captures how research is conducted at universities in countries such as Austria, France, Ireland, Ukraine and the UK, it misses the more hierarchical structure of, say, German universities, where research groups are more focused around individual senior professors.}  
Two such plots are given in Fig.~1: one for Physics and one for Applied Mathematics. 
Almost all experimental physics submitted to the RAE were assigned to the ``Physics'' UoA. Some theoretical  also came under the ``Physics'' banner but others were entered into ``Applied  Mathematics''.
For our purposes then, we may consider ``Physics'' as meaning experimental physics and ``Applied Mathematics'' as being both applied mathematics and theoretical physics.
Fitting to the formula (\ref{seven}) delivered the estimates $N_c = 13 \pm 2$
for Applied Mathematics and $N_c = 25 \pm 5$ for physics. 
Note how, as expected, the slopes of the $s$ versus $N$ curves reduce to the right of the Dunbar number. 
Note also that the right slope is smaller for the discipline with the higher $N_c$-value, as expected.

\begin{table}[b!]
\caption{The values of the Dunbar numbers or upper critical masses  for a variety of academic disciplines.}
\begin{center}
\begin{tabular}{|l|r|} \hline \hline
                                         &                \\
Research discipline                                & $N_c\quad\quad$                 \\
                                                   &                            \\
\hline
Applied mathematics                                &$13 \pm  2~\,$     \\
Statistics \& operational research                 &$17 \pm  6~\,$     \\
Physics                                            &$25 \pm  5~\,$     \\
Geography, environment \& Earth studies            &$30 \pm  3~\,$     \\
Biology                                            &$21 \pm  4~\,$     \\
Chemistry                                          &$36 \pm  13$     \\
Agriculture, veterinary \& food sciences           &$10 \pm  3~\,$     \\
Law                                                &$31 \pm  4~\,$     \\
Architecture, the build environment,
town \& country planning                           &$14 \pm  3~\,$     \\
French, German, Dutch \& Scandinavian 
languages                                          &$ 6 \pm  1~\,$     \\
English language \& literature                     &$32 \pm  3~\,$     \\
Pure mathematics                                   &$ \le 4   ~\,~\,~\,~\,$     \\
Medical sciences                                   &$41 \pm  8~\, $     \\
Nursing, midwifery, allied health
professions \& studies                             &$18 \pm  5~\, $     \\
Computer sciences 1                                &$11 \pm  5~\,$ \\
Computer sciences 2                                &$33 \pm  9~\,$  \\
Computer sciences 3                                &$49 \pm 10$ \\
Archaeology  1                                     &$17 \pm  3~\,$     \\
Archaeology 2                                      &$25 \pm  4~\,$  \\
Economics and econometrics                         &$11 \pm  3~\,$     \\
Business and management studies                    &$48 \pm  8~\,$     \\
Politics and international studies                 &$25 \pm  5~\,$     \\
Sociology                                          &$14 \pm  4~\,$     \\
Education                                          &$29 \pm  5~\,$     \\
History                                            &$25 \pm  5~\,$     \\
Philosophy and theology                            &$19 \pm  3~\,$     \\
Art \& design                                      &$25 \pm  8~\,$     \\
History of art, performing  arts, 
communication and music                            &$ 9 \pm  2~\,$     \\
                                          &                \\
\hline \hline
\end{tabular}
\end{center}
\end{table}

We have produced similar plots to those in Fig.1 for a variety of academic disciplines based on data from 
RAE2008. The reader is referred to the original literature for details~\cite{KeBe11a}.
We reproduce the $N_c$ values in Table~1 for convenience.
Three different candidate Dunbar numbers were detected for Computer Science. We speculate that this discipline comprises a number of very different approaches, from the formal and theoretical to the empirical. 
Similarly in Archaeology, two candidates for $N_c$ were determined.
In Pure Mathematics no breakpoint was found and since the smallest group submitted had $N=4$, we believe the $N_c$ value for that discipline is that number or less. 

We have stated that $N_c$ is the Dunbar number or upper critical mass. 
However, this is {\emph{not} analogous the old (unfounded) notions of critical mass as a threshold
below which high quality research is impossible. 
(Clearly there is no such threshold.)
Nonetheless, we can come closer to the old concept of critical mass~\cite{Ha09} with the following considerations.

Suppose funding for a new academic were to become available.
We ask whether it is more beneficial for society as a whole  to allocate the new researcher  
to a group with $N>N_c$ or to one with $N<N_c$. 
The decision is made by considering the rate of change of {\emph{strength}} $S=sN$ with $N$.
We find that $dS/dN$ is bigger for $N<N_c$ groups provided that $N>N_c/2$.
In other words, to maximally benefit society as a whole, additional researchers should be given to the smaller groups provided they are not too small. The cutoff is at 
\begin{equation}
 N_k = \frac{N_c}{2}.
\label{Nk}
\end{equation}
We refer to this number as the {\emph{critical mass}}. 
(In our earlier papers we called it {\emph{lower critical mass}}.
In other words, we have established a scaling relation between the critical mass and the Dunbar number.
 
We can now divide research groups into three different categories: small ($N < N_k$); medium ($N_k < N < N_c$); and large ($N>N_c$).
These categories are discipline-dependent.
 For example, while a team of five pure mathematicians
may be considered as ``large'', a similar number of experimental physicists is ``small''.
This accords with experience.

The overall shapes of the curves in Fig.1 may be ascribed to a Ringelmann-type effect.
The original Ringelmann effect, in psychology and in sports science, is the tendency for individual members of a group to become less productive as the  group size increases~\cite{Ri13}. 
The effect is due to an increase in  inefficiency as the numbers grow. 
In psychology, this is often ascribed to motivational losses (also known as social loafing).
But coordination losses may also account for the  phenomenon.
Here, as in Dunbar's theory, we also believe that the phenomenon is due to coordination losses.
However, rather than the decrease in the average individual performance
(which would be the actual Ringelmann effect), our model manifests a reduction in the {\emph{rate of change of quality}} per unit staff member. 
In other words, it is the quality-versus-quantity gradient rather than
in the quality itself that reduces as $N$ increases. 
The challenge to managers, therefore, is to counteract or minimise this effect. 
Can we find point the direction to looking for a solution? We believe we can do so as follows.

The dotted line in Fig.1 is an extrapolation of the left fit for small and medium groups into the supercritical region where $N$ exceeds $N_c$. 
We refer to such extrapolated lines as ``overshoots''. 
In the absence of a transition point $N_c$, if the mechanism which governs research quality for  small and medium universities applied also to the large ones, then the research quality to the right of $N_c$ would also be expected to follow this line.
In other words, the naive model outlined earlier, and the one upon which many critical mass policies are currently based, would predict an unbounded Matthew effect along this line. 
In this case, a policy of continued concentration of
resources would indeed drive up research quality. 
However, the figure clearly indicates that this is not the case.
Similar figures for other research disciplines may be found in Ref.\cite{KeBe12a}. 

Instead, as the figure clearly demonstrates,   large
research teams tend to have a different interaction pattern than small and medium ones, as predicted in Eq.(\ref{Nc}). 
With a large
value of $N_c \approx 25$ for Physics, research quality is saturated to the right of the $N_c$ as Dunbar's theory predicts. 
This suggests that  the concentration of more staff into these teams only leads to a linear increase in
research volume and \emph{not} to an increase in research quality.

However, it is interesting to note that some of the best performing research teams, which appear as outliers to the overall fit, are bunched slightly above $N_c$ but are well described by the overshoot. 
We suggest that this overshoot may be
caused by a greater than usual degree of cohesiveness in these highly successful research teams.
Somehow, two-way communication links are sustained despite their group sizes exceeding the Dunbar numbers for their disciplines.
These are the research groups which are best managed: those in which communication is key.

We have shown that, it is most beneficial to society if new researchers are added to medium-sized groups.
This is because the new researcher brings new links to existing academics in such groups, helping to drive them to the Dunbar number. 
However, joining a medium-sized group is {\emph{not}} the best course of action from the new individual's point of view.
For them, it is better to join a large group. 
The reason is that a medium group (unless it is on the border of becoming large) only provides the new researcher with a number of links which is below their cognitive limit.
If they join a large group, on the other hand, they can enjoy maximal connectivity.
(Of course establishing such connections takes time as the intra-group cooperative network becomes reconfigured.)
This returns us to the causation mechanism mentioned at the start of Section~2. 
Of course high-quality individual researchers are attracted to high-quality, large groups. 
However, that is not how quality grows.
Adding more and more researchers to large groups only increases the volume of high-quality outputs.
It cannot increase average quality significantly because that is already at saturation level.
The fastest way to drive up research quality is to help medium-sized groups attain the nirvana that is the Dunbar number. 

\section{Quality of Evaluation: Quantitative Indicators}
\label{III}
\setcounter{equation}{0}

So far, we addressed evaluation of the quality of research and the roles played by Dunbar's numbers and critical masses. 
Now we turn to the second main theme announced in the title of our paper: 
the quality of research evaluation through metrics.

Nowadays, many different measures exist claiming to assess academic research excellence or impact
and these are subject to ongoing discussion and debate within the academic, scientometric,
university-management and policy-making communities internationally.
A topic of prime importance is the extent to which citation-based indicators compare
with peer-review-based evaluation.
Although flawed in many obvious respects, peer-review remains the only approach that is broadly accepted by the scientific community. 
But peer-review exercises such as the RAE and REF are expensive, both in terms of the labour involved and the time lost by academics who could otherwise have been engaged in research activity.
For example, it has been estimated that UK universities spent about $\pounds$4000 for each researcher submitted to REF2014~\cite{cost}.
The total cost to the UK of running the national evaluation exercise in 2014 is estimated to have been  $\pounds$246 million~\cite{cost}.
This is a significant fraction of the $\pounds$1.6 billion distributed by HEFCE for quality-related research  funding in 2015-2016. 
(That figure is based on submission of 52\,077 individuals, so the funding allocated to universities is over  $\pounds$30,000 per submitted researcher in that year.)

For this reason, it is desirable for governments, funders and policy makers  to introduce, if it exists, a simple, cheap and reliable way to measure  scientific excellence.
In the absence of expert subject knowledge, professional research managers are also keen on such an approach to serve as a basis for regularly measuring the performances of research groups and individuals.
However, while citation-based indicators and metrics are readily accessible, they are far from being universally accepted as a way to inform evaluation processes.
They are even less accepted by academics as a way to replace evaluations based
on peer review such as the RAE and REF. 

In Refs.\cite{MrKe13a,MrKe13b}, we considered  a citation-based measurement of research at an
amalgamated, institutional or group level, from the natural to social sciences and humanities.
In particular, we examined how that measure correlates with the results of RAE2008.
We found that the citation-based indicator is very highly
correlated with peer-evaluated measures of group strength $S$ for some disciplines.
But it is poorly correlated with group quality $s = S/N$ for all disciplines. 
This means that,  almost paradoxically, indicators
could possibly form a basis for deciding on how to fund research institutions, especially for the so-called hard sciences, but they should not be used as a basis for ranking  or comparison of research groups. Moreover, the correlation between peer-evaluated and citation-based scores is weaker
for soft sciences.

The citation-based measure which we compared against peer review in Refs.\cite{MrKe13a,MrKe13b} was one  provided by {\emph{Thomson Reuters Research Analytics.}} 
This company has developed the so-called {\emph{normalised citation impact}} (NCI) $i$ as a measure of a department's citation performance in a given discipline.
The NCI is calculated using data from the {\emph{Web of Knowledge} databases and by comparing outcomes to a mean or expected citation rate. 
The measure is determined for an entire group or department and then normalised by group size. 
It is therefore a {\emph{specific}} (per-head) measure. 
A useful feature of  the NCI is that it attempts to take account of
different citation rates in different  disciplines. 
To achieve this,  the total citation count for each paper is  ``rebased'' to an average number
of citations per paper for the year of publication and either the field or journal
in which the paper was published. 
Here we denote the NCI specific measure by $i$. 
Scaled up to the size of a group or department, the
corresponding {\emph{absolute}} measure   is denoted by $I$ where $I = iN$.

The relationship between $i$ and $s$ for Physics research groups submitted to RAE2008 is given in Fig.2 (left panel).
Clearly the correlation between the two measures is very weak.
In fact the Pearson correlation coefficient is only $0.48$. 
In Refs.\cite{MrKe13a,MrKe13b} we also analysed the correspondence between $i$ and $s$ for 
biology; chemistry; mechanical, aeronautical and manufacturing engineering; geography and environmental studies; sociology; and history. 
The maximum value for Pearson's coefficient was 0.60 (for biology and chemistry). 
These weak correlations indicate that the NCI cannot be use to estimate research quality - 
it cannot be used as a proxy for peer review.

We also ranked the various universities according to their $i$ and $s$ values to see if the NCI could, at least, serve as a guide to league tables. However Spearman's coefficient was of similar  weakness  to  Pearson's so this is also ruled out.

\begin{figure*}[t]
\begin{center}
\includegraphics[width=0.47\columnwidth,angle=0]{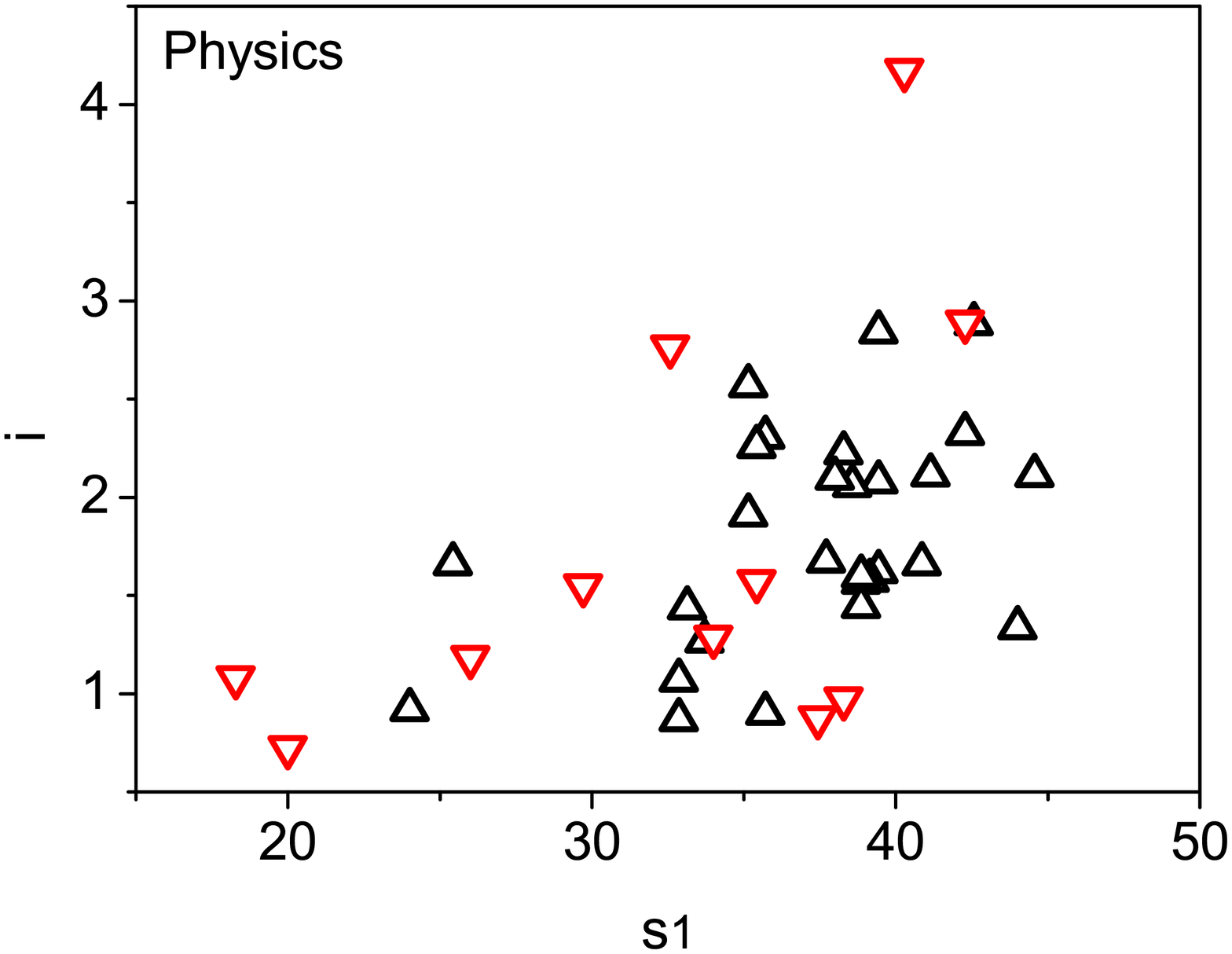}
\includegraphics[width=0.47\columnwidth,angle=0]{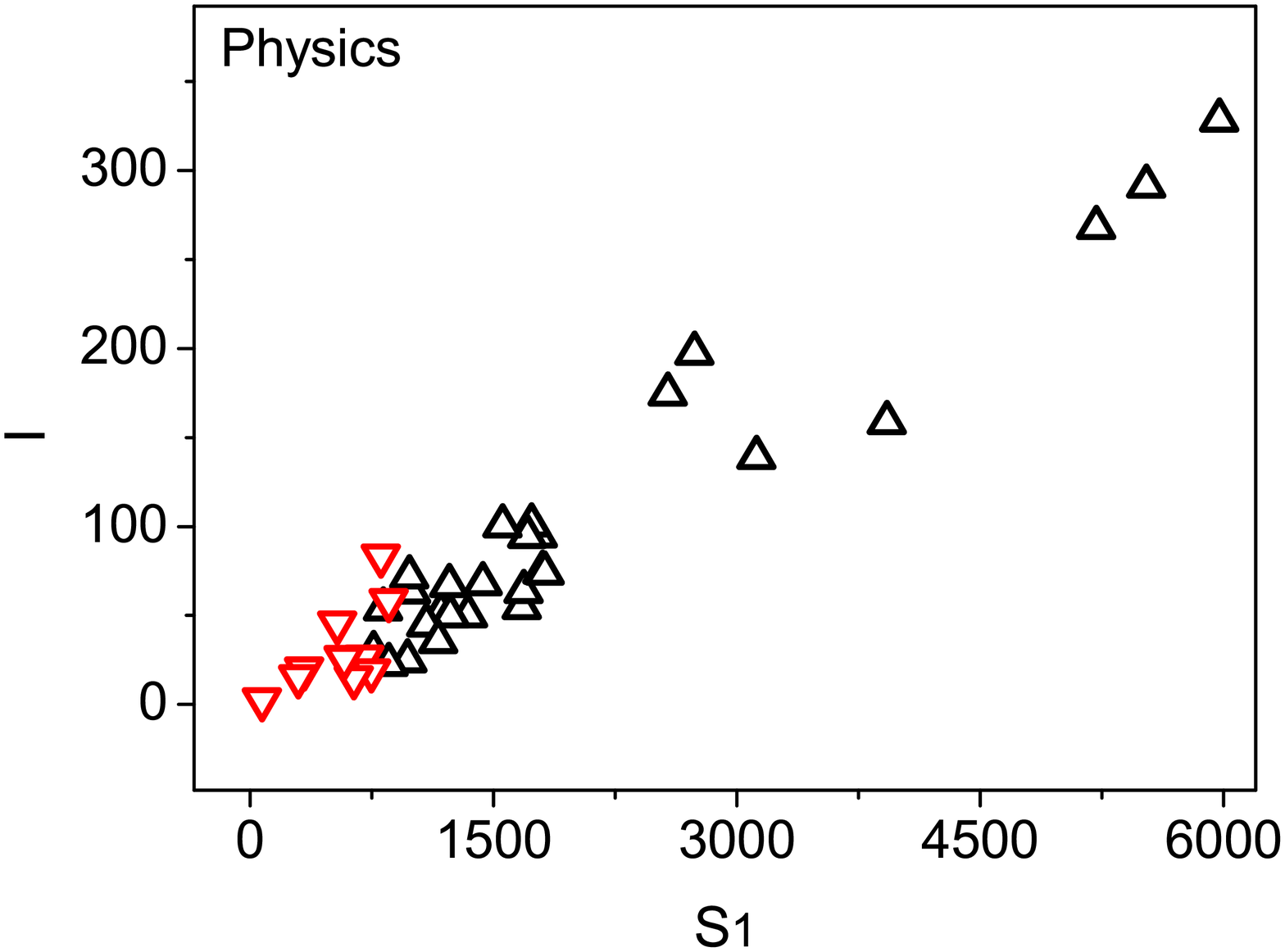}
\label{fig2}
\end{center}
\vspace{0.3cm}
\caption{Left panel: The correlation between the normalised citation impact $i$ and peer-review measures of group research quality $s$ for Physics.
Right panel: The correlation between the absolute  impact $I=iN$ and  research strength/power $S=sN$.
 The black and red  symbols represent large and medium/small  groups respectively.}
\end{figure*}

Figure~2 (right panel) depicts the relationship between the absolute measure $I = iN$ and strength (or power) $S = sN$. 
The correlations are very good. 
In fact, for physics, Pearson's correlation coefficient for these measures is a remarkable 0.96. 
It is also in the nineties for the other scientific disciplines (biology and chemistry). However it is lower for sociology and history (0.88 in each case). 
The multiplication by $N$ when replacing  specific measures by their absolute counterparts stretches 
the data sets and the correlations between the rescaling  leads to improved correlations.
Moreover, when one partitions the various teams into small; medium; and large (as explained in the previous section), the correlation 
tends to be best for large groups.
Nonetheless the conclusion is clear: citation-based measures may perhaps
inform,  or even serve as a proxy,  for peer-review  measures of the strengths of research
groups. But they cannot be used for measures of quality.

In the UK system, quality-related governmental funding for research is strength based; 
the amount of money delivered to universities for a particular research discipline is a function of their $S$-scores at RAE or REF. 
The above results indicate that the use of citation-based alternatives may offer a far cheaper, and less intrusive alternative to the current system.
The scale of the savings that could be made in the UK alone is  a quarter of a {billion} pounds per exercise. 
Such a proxy would work best for the hard sciences but would be less reliable for the social
sciences and humanities. 
Also, it is crucial to realise that such citation-based indicators should not be used
 to compare or rank the qualities of research groups.

However, we see a grave danger in introducing any form of citation-based metric to evaluation exercises.
The {\emph{Hawthorne effect}} (also known as the {\emph{observer effect}}) is the name given to reactions of  individuals when an aspect of their behavior is observed or measured; 
they naturally seek to modify that behaviour. 
This would be disastrous in the academic context; university managers would force researchers to chase citations instead of curiosity.
This would shift the entire landscape of academic research as everyone chases the same ``hot topics'' in a bid to increase their citation counts.
It is well known that many profound and important scientific breakthroughs have come about through mixtures of curiosity, serendipity, accident and chance. 
Academic freedom is the cornerstone on which these features can thrive.
Combined with the professionalisation of research management and ``businessification'' (indeed, moves towards privatisation) of many third-level institutions, citation-based league tables would pose a serious threat to academic freedom.

In 2014, Dorothy Bishop arrived at a similar conclusion to us. 
She used a different metric - the so-called {\emph{departmental $h$-index} instead of the NCI.
In Ref.\cite{MrKe15a} we showed that Bishop's metric has an even better correlation with 
the RAE-measured strength index $S$ than has the scaled-up NCI. 
The meaning of a departmental $h$-index of, say, $n$ is that $n$ papers, authored by
researchers from a given department  in a given discipline  were cited at least $n$ times over a
given time period (we used the periods of assesment for RAE2008 and REF2014). 
This index takes into account, in principle,  {\emph{all}} researchers (not only those submitted to RAE or REF). However, in practice it can be dominated by a single, extremely strong individual.
Also, unlike for REF where authors' addresses at the census date determine their affiliations for assessment purposes,  author address at the time of publication determines to which university a given
output is allocated for the departmental $h$-index.

Nonetheless, the departmental $h$-index is indeed better correlated with research quality; the Pearson and Spearman correlation coefficients for between $h$ and $s$ for Physics at RAE2008 were  0.55 and 0.58, respectively. 

These correlations are still, however, too weak to replace or inform exercises like the RAE or REF.
To demonstrate that, we decided to use the departmental $h$-index to predict outcomes, in terms of rankings, of RAE2014 in advance of their being announced by HEFCE.
If simple citation-based metrics can be used as some sort of proxy for  peer 
review, one would expect them to be able to predict  at least some such
aspects of the outcomes  of such exercises. 
Even limited success might suggest that a citation metric could serve at least as a ``navigator'' -- to help guide  research institutes as they prepare for the expert exercises.

We placed our predictions for the rankings in Biology, Chemistry, Physics and Sociology on the arXiv in November 2014
(they were subsequently published as Ref.\cite{MrKe15a}.)
The REF results were announced in December 2014.
We revisited our  study in January 2015 and found that our
predictions failed to anticipate REF outcomes.
For example, the Pearson and Spearman correlation coefficients between the REF-measured $s$ values and the departmental $h$-index in Physics were 0.55 and 0.50, respectively.
One submission which was ranked 27th in a certain subject area according to the departmental $h$-index, actually came in in seventh place in the REF.

We also sought to predict whether institutions would move up or down in the rankings between RAE2008 and REF2014.
For Physics, the correlation between our predictions and the actual results was 0.26.
For Chemistry it was 0.05 and for Biology it was -0.15.
As commented later, managers would find better estimates of movement in the league tables by tossing dice.
Our results are published as Ref.\cite{MrKe15b}.

\section{Conclusions}
\label{IV}
\setcounter{equation}{0}

In the first part of this review, we reported upon the development of a precise definition of a  minimum {\emph{critical mass}} for academic research. 
(In our earlier publications on the topic, we referred to this quantity as ``lower critical mass''.)
Contrary to previous intuition-based notions, this is not a threshold value marking a step change between low- and high-quality research. 
In fact it is not directly perceptible or measurable.
Instead it is connected, via a scaling relation, to the  Dunbar number for academic research. 
(We previously referred to this as the ``upper critical mass''.)
This marks an average cognitive limit to the number of intra-group research relationships one can sustain; in a large department, one cannot cooperate with everybody all of the time.
The Dunbar number is discipline dependent; e.g., it is less than or equal to four for Pure Mathematics and approximately twenty-five for Experimental Physics on average (with a standard deviation of five).
 
We showed that concentrating more staffing resources into already-large research groups does not lift the quality of those groups significantly, especially if the discipline has a high Dunbar number. 
This is because everyone already has sufficient links to maximise cognitive activity. 
Adding more resources to such a group only serves to increase the volume of high-quality research.
If one wishes to increase quality more significantly, more research staff should be allocated to medium-sized research groups.  
The standard is raised by introducing  new links, as relationships between existing staff and the new arrival are established over time. 

We noticed in Fig.1, and similar figures for other disciplines, that the best performing groups are not always the biggest. 
Frequently groups with size just in excess of the Dunbar number perform exceptionally well - they are large groups behaving as if they were medium.
For example, in REF2008, Lancaster University with 26 Physics staff submitted, was ranked highest in terms of quality but Cambridge, with 141.25 staff, was largest.
 We suggested that these are the groups within which communication flows easiest and we suggested that these should form a guide for managers at other institutions to seek to emulate.

We then compared measures of quality of research coming from peer review to a citation-based measures.
The poor correlations  mean that such citation indicators {\emph{cannot}} be used as proxies for peer.
Nor should such indicators be used to rank university departments or groups into league tables.

However, the NCI and $h$-index compare well to the outcomes of peer review when one is interested in measuring the strength or power, defined as volume of quality.
Since this is the basis on which quality-related funding is distributed, one could argue that a suitable metric may be developed to replace invasive exercises or  reduce their cost.
Dorothy Bishop goes a step further; she suggests to  abolish the REF altogether and revert to a block-grant system that existed before the RAE.
She argues that besides not being  cost-effective, the REF has numerous adverse effects on academia.

On the basis of  evidence and rigorous analyses produced over the past five years, we are inclined to agree, but only for well-established universities with large research groups of solid and long-standing repute. 
Research evaluation exercises have, however, been useful for smaller and medium groups as they seek to establish themselves in newer universities which have less developed research cultures.
Some such universities have learned -- through the RAE and REF -- that curiosity-driven, pure academic research has value.
We suggest that peer review (not metric-based!) exercises for such groups can and should continue.
But they should be extended to include measurements of the extent to which such institutes promote academic values such as academic freedom and bottom-up research.
In other words, peer review should be more focused on institutions which can actually benefit from it.
Those which are established, and who enjoy long-standing high reputations amongst their peers internationally, should be allowed to get on with research.
Fig.1(a) shows that very little separates such groups in terms of quality; in ranking such groups, one is essentially ranking  noise. 

At the risk of stating the obvious, citation-based metrics measure citation counts, not quality.
These are very different things. 
We suggest that on no account should one use metrics to decide policy or inform management.
In our opinions, these pose a serious and current threat to curiosity-driven research which is the very {\emph{raison d'\^{e}tre}} for universities.

Finally we suggest that it is incumbent upon academics themselves, who have the training and expert knowledge, to turn their methodologies on the scientific process itself in order to produce evidence to influence the powers that be. 
The UK recently commissioned an independent review of the role of metrics in research assessment and management.  
The remit was ``to investigate the current and potential future roles that quantitative indicators can play in the assessment and management of research.'' 
The report, ``The Metric Tide'', was published in July 2015~\cite{Tide1}.
It concludes that  peer review continues to command widespread support as the main basis for evaluating research and there is legitimate concern that  indicators, including citation counts and rankings, can be misused or ``gamed.''
The report also  concluded  that no current metric can  provide a replacement for  peer review.

A  legacy of the Report is the establishment of a  forum for ongoing discussion of  issues related to metrics.
This aims to celebrate good practices but also to name and shame bad ones. 
Inspired by the ``Bad Sex in Fiction'' award, a  ``Bad Metric'' prize will be awarded to the most egregious example of inappropriate use of quantitative indicators in research management~\cite{ResponsibleMetrics}.
This demonstrates the power of  turning the evaluation process on itself





\section*{References}
\begin{thebibliography}{9}


\bibitem{KeBe10}
Kenna R and Berche B 2010
EPL {\bf{90}} 58002.

\bibitem{KeBe11a}
Kenna R and Berche B 2011
Scientometrics {\bf{86}} 527 - 540.




\bibitem{KeBe12a}
Kenna R and Berche B 2012
IMA Journal of Management Mathematics {\bf{23}} 195-207.



\bibitem{Bi14}
Bishop D 2014 
http://deevybee.blogspot.co.at/2013/01/an-alternative-to-ref2014.html.
Accessed 01 November 2015.



\bibitem{Evidenceweb}
Evidence Thomson Reuters 
http://www.evidence.co.uk. 
Accessed 01 November 2015.

\bibitem{Ev10} 
Evidence Thomson Reuters 2010 
{\emph{The future of the UK University research base}}
http://www.universitiesuk.ac.uk/Publications/Documents/UUK-FutureOfResearch-LiteratureReview.pdf.

\bibitem{Evidence2011}
Evidence Thomson Reuters 2011
{\emph{Funding research excellence: research group size, critical
mass \& performance}}.





\bibitem{Ha09}
Harrison M 2009 
in The {\emph{Question of R\&D Specialisation: Perspectives and Policy Implications,}} 
edited by D Pontikakis, D Kriakou and R van Baval, JRC Technical and Scientific Reports (European Commission, Luxembourg), 57-59.
	


\bibitem{Nature}
Davide Castelvecchi 2015 
Nature {\bf{256}}  179-181.



\bibitem{REF2014def}
{\emph{REF 2014 rerun: who are the 'game players'?}},
https://www.timeshighereducation.com/features/ref-2014-rerun-who-are-the-game-players/2017670.article.
Accessed 01 november 2015.


\bibitem{Du92}
Dunbar R I M  (1992)
Journal of Human Evolution {\bf{22}} 469–493.



\bibitem{Ri13}
Ringelmann M 1913
Annales de l'Institut National Agronomique {\bf{2}} 2-39.




\bibitem{cost}
Farla K and Simmonds P (2015)
{\emph{REF Accountability Review: Costs, benefits and burden}}
Report by Technopolis to the four UK higher education funding bodies
http://www.hefce.ac.uk/pubs/rereports/Year/2015/refreviewcosts/Title,104406,en.html.
Accessed 31.10.2015.


\bibitem{MrKe13a}
Mryglod O, Kenna R, Holovatch Yu and Berche B 2013
Scientometrics {\bf{95}} 115-127.

\bibitem{MrKe13b}
Mryglod O, Kenna R, Holovatch Yu and Berche B 2013
Scientometrics {\bf{97}} 767-777.


\bibitem{MrKe15a}
Mryglod O, Kenna R, Holovatch Yu and Berche B 2015
Scientometrics {\bf{102}} 2165-2180.

\bibitem{MrKe15b}
Mryglod O, Kenna R, Holovatch Yu and Berche B 2015
Scientometrics {\bf{104}} 1013-1017. 



\bibitem{Tide1}
The Metric Tide: Report of the Independent Review of the Role of Metrics in Research Assessment and Management (July 2015)
http://www.hefce.ac.uk/pubs/rereports/Year/2015/metrictide/Title,104463,en.html.
Accessed 01 November 2015.



\bibitem{ResponsibleMetrics}
Responsible Metrics
www.ResponsibleMetrics.org.
Accessed 01 November 2015.








\end{thebibliography}
\end{document}